\documentclass[preprint]{aastex}
\usepackage{emulateapj5,apjfonts,psfig}

\newcommand{\Msun}{M_\odot} 
\newcommand{\kms}{$\rm {km}~\rm s^{-1}$}

\slugcomment{{\it The Astrophysical Journal}.}
\righthead{BLACK HOLE IN G1}
\lefthead{GEBHARDT, RICH, \& HO}

\begin{document}
 
\title{An Intermediate-Mass Black Hole in the Globular Cluster G1:\\
Improved Significance from New Keck and Hubble Space Telescope
Observations\footnotemark[1]}

\footnotetext[1]{Based on observations made with the {\it Hubble Space
Telescope}, which is operated by AURA, Inc., under NASA contract
NAS5-26555.}
 
\author{Karl Gebhardt\altaffilmark{2}, R. M. Rich\altaffilmark{3}, and 
Luis C. Ho\altaffilmark{4}}

\altaffiltext{2}{Astronomy Department, University of Texas, Austin, TX
78723; gebhardt@astro.as.utexas.edu}

\altaffiltext{3}{UCLA, Physics and Astronomy Department, Math-Sciences 8979,
Los Angeles CA 90095-1562; rmr@astro.ucla.edu}

\altaffiltext{4}{The Observatories of the Carnegie Institution of
Washington, 813 Santa Barbara St., Pasadena, CA 91101; lho@ociw.edu}

\begin{abstract}

We present dynamical models for the massive globular cluster G1. The
goal is to measure or place a significant upper limit on the mass of
any central black hole. Whether or not globular clusters contain
central massive black holes has important consequences for a variety
of studies. We use new kinematic data obtained with Keck and new
photometry from the Hubble Space Telescope. The Keck spectra allow us
to obtain kinematics out to large radii that are required to pin down
the mass-to-light ratio of the dynamical model and the orbital
structure. The Hubble Space Telescope observations give us a factor of
two better spatial resolution for the surface brightness profile. By
fitting non-parametric, spherical, isotropic models we find a best-fit
black hole mass of $1.7(\pm0.3)\times10^4~\Msun$. Fully general
axisymmetric orbit-based models give similar results, with a black
hole mass of $1.8(\pm0.5)\times10^4~\Msun$. The no-black hole model
has $\Delta\chi^2=5$ (marginalized over mass-to-light ratio), implying
less than 3\% significance. We have taken into account any change in
the mass-to-light ratio in the center due to stellar remnants. These
results are consistent with our previous estimate in Gebhardt, Rich \&
Ho (2002), and inconsistent with the analysis of Baumgardt et
al. (2003) who claim that G1 does not show evidence for a black
hole. These new results make G1 the best example of a cluster that
contains an intermediate-mass black hole.

\end{abstract}


\section{Introduction}

A fundamental problem in understanding galaxy formation is knowing how
supermassive black holes form and grow. The correlation between black
holes and host galaxy properties (with velocity dispersion: Gebhardt
et al. 2000a, 2000b; Ferrarese \& Merritt 2000; or with bulge mass:
Magorrian et al. 1998) highlights an intimate connection. Theoretical
models are beginning to explain these correlations in detail and
suggest that the supermassive black hole has significant long-ranging
influence on the host galaxy (Silk \& Rees 1998, Fabian 1999,
Springel, Di Matteo, \& Hernquist 2005, Murray, Quataert, \& Thompson
2005, etc.). However, one of the main issues is understanding the
seeds for the supermassive black holes, since they determine the
initial mass and growth process; i.e., if intermediate-mass black
holes are common then there is likely to be a significant number of
merging events. Thus, the existence and number density of
intermediate-mass black holes is one of the most important pieces to
the puzzle.

Recent detections of active galactic nuclei (AGNs) in low-luminosity,
late-type galaxies strongly suggest that intermediate-mass black holes
do exist (Filippenko \& Ho 2003; Barth et al. 2004; Greene \& Ho
2004).  Combining black hole masses estimated using the AGN luminosity
and line width with stellar velocity dispersions of the host galaxies,
Barth, Greene, \& Ho (2005) demonstrate that the black hole
mass-velocity dispersion relation extends down to $\sim 10^5$
$M_\odot$.  Intermediate-mass black holes have also been invoked to
explain the origin of the ``ultraluminous X-ray sources'' detected in
nearby galaxies by Chandra (e.g., Zezas \& Fabbiano 2002, Kong et
al. 2005) and XMM-Newton (e.g., Foschini et al. 2002), but these
measures depend on uncertain models for accretion disks, and
stellar-mass black hole models may fit as well (King et al. 2001). The
most robust way to measure black hole masses is through dynamics. The
two claims for dynamical evidence for intermediate-mass black holes
are for G1 in M31 (Gebhardt, Rich, \& Ho 2002) and M15 (van der Marel
et al. 2002; Gerssen et al. 2002), both of which have been challenged
by Baumgardt et al. (2003a,b). Here we present analysis of G1 and
de~Zeeuw et al. (2005) present new analysis for M15.

In Gebhardt, Rich \& Ho (2002), we present data and dynamical models
for G1 that suggest the existence of a $2\times10^4~\Msun$ black hole.
Subsequently, Baumgardt et al. (2003b) compare our data to their
dynamical models and argue that a model with no black hole fits as
well. In this paper, new data and analysis strongly support the black
hole interpretation. Furthermore, we argue that the Baumgardt analysis
is fundamentally inconclusive since they use a simplified comparison
between data and theory. Thus, even with the data as presented in
Gebhardt, Rich \& Ho, the black hole interpretation is preferred, but
the new data presented here give yet stronger evidence.

First we outline the appropriate techniques that should be used to
measure black hole masses, and discuss why our previous analysis and
modeling are preferred. We then provide new data and results which
strengthen the black hole interpretation.

\section{Measuring Black Hole Masses}

The history of supermassive black hole studies is the subject of many
reviews (e.g., Kormendy \& Richstone 1995, Kormendy \& Gebhardt 2001,
Ho 1999). Here we concentrate on the main kinematic requirements for
accurate measurements and the modeling techniques for optimal
constraints of central black holes. It has been know since Binney \&
Mamon (1982) and Tonry (1983) that using the second moment of the
velocity profile alone to measure the mass profile can lead to
substantially biased results. The problem is that the velocity
anisotropies of the stars trade off with the shape of the potential to
create a variety of profiles for the second moment. Thus, any
assumption about the stellar orbital distribution must be explored
carefully. One, however, can get a handle on the orbital distribution
by exploiting at least the first moment of the velocity distribution
(i.e., the velocity) along with the velocity dispersion, and, more
importantly, higher-order terms. Van der Marel (1991) demonstrates the
power of using additional information to extract the orbital
structure. The most information that can be extracted from any
spectral dataset results from using the full line-of-sight velocity
distribution (LOSVD) at as many positions in the object as
possible. Using velocity dispersion data alone (or even the second
moment) cannot overcome the degeneracy with the orbital
structure. Magorrian et al. (1998) demonstrate how sensitive the
estimate of the black hole mass is to changes in the assumed
anisotropy when using the second moment alone. The current state of
the art is to use the full velocity profile when possible.

The dynamical modeling is just as important as the data analysis,
since any assumptions in the models will greatly bias the orbital
structure and hence the mass profile. Gebhardt (2004) reviews the
classes of models that have been used to measure the central
potential. One must use orbit-based models in order to include
degeneracies with the stellar orbital structure. These are now
commonplace and standard (Gebhardt et al. 2000c, 2003; Cappellari et
al. 2003; Krajnovic et al. 2005). These models provide the most
freedom for the distribution function in axisymmetric systems
(triaxial models are now in development---see van de Ven et
al. 2004---and will soon be as common). Alternatively, N-body
simulations provide just as general results in terms of priors on the
distribution function. N-body simulations that reach the size of
realistic clusters and galactic nuclei will therefore be another tool
for the study of central black holes.  Gebhardt, Rich \& Ho (2003) use
orbit-based models and Baumgardt et al. (2003b) use N-body simulations
to constrain the central black hole mass. The difference in the two
results is due to {\it both} how they include the kinematics and the
theoretical comparison. We discuss both of these.

\subsection{Velocity Dispersion is Not Enough}

First, Gebhardt, Rich \& Ho model the full velocity profile (as is
done in this paper) whereas Baumgardt et al. only use the second
moment (they actually plot the first moment but never use it in their
$\chi^2$ measurements). It is known that the second moment alone is
not adequate in general. However, for an object like G1 where the
sphere of influence of the black hole is barely resolved, it is
crucial to use all of the kinematic information. Specifically, the
signature of the black hole is one of increasing the wings of the
velocity profile near the center. The extreme consequence is to cause
exponential tails in the central velocity profile, but this would only
be seen in the best spatially resolved cases (as demonstrated for M87
by van der Marel 1994). For G1, if Gebhardt, Rich \& Ho had used the
second moment alone to constrain the black hole mass they, too, would
have argued that the no-black hole model is acceptable. In fact, the
no-black hole model of Gebhardt, Rich \& Ho provides a better fit to
the second moment than the no-black hole model of Baumgardt et
al. Yet, ironically, Baumgardt et al. argue for no black hole and we
argue for a black hole.  The difference is because we use more
information than used by Baumgardt et al. Unfortunately, the N-body
simulations do not have enough particles to measure a reliable
velocity profile. So the only recourse that Baumgardt et al. have is
to use the second moment (even the first moment is difficult to
measure due to the shot noise in the N-body simulations). This leaves
us in the unfortunate situation where we cannot use the same dataset
for comparison. However, we strongly argue that by not using the full
velocity profile one is severely limiting the generality of the
results and potentially introducing biases.

\subsection{Parameter Estimation vs. Hypothesis Testing}

Second, Gebhardt, Rich \& Ho compare black hole and no-black hole
models in a differential sense, whereas Baumgardt et al. use an
absolute comparison between their no-black hole model and our data. In
order to determine whether inclusion of a black hole provides a better
fit, one must use a differential comparison. This issue is discussed
in Gebhardt et al. (2000c), but we summarize the main points here. The
two analyses highlight the difference between parameter estimation
(our analysis) and hypothesis testing (Baumgardt et
al. analysis). Hypothesis testing is the most base level analysis that
has very little power to discriminate whether a black hole exists. For
example, if a kinematic dataset has a significant amount of data
outside of the black hole sphere of influence, the majority of a
goodness-of-fit statistic is dominated by regions that cannot
discriminate whether a black hole is present. Thus, even if a central
kinematic measurement is not in agreement, the hypothesis test will
de-weight that sole deviation. In parameter estimation, a difference
test (i.e., using $\Delta\chi^2$), directly shows the effect of
including a black hole, and is therefore not sensitive to the amount
of data at large radii. Since the black hole in G1 has such a small
sphere of influence it is imperative to use a differential
analysis. The main problem with the Baumgardt et al. analysis is that
they cannot run realistic N-body simulations that include a black hole
(they have too few particles), so they are forced to perform a
hypothesis test. Thus, their result is severely compromised to the
point where one can draw very little meaning to their comparison to
the G1 data.

\subsection{Other Issues}

Above are the two main concerns, but additional worries about the
N-body simulations exist. Other issues are the scaling of results to
realistic clusters and their initial hidden assumptions. For G1, even
for the largest N-body simulation, they must scale the mass by a
factor of over 150. It has not been demonstrated that such an extreme
scaling produces accurate results, especially near the cluster
center. In fact, Baumgardt (2001) demonstrate how difficult it is to
scale even simple single-mass models. For this resolution, we have to
wait until the simulations become more sophisticated. Another issue
with the N-body simulations is the influence of binary
stars. Baumgardt et al. also do not include binary stars, which are
likely to dominate the core dynamics (Fregeau et al. 2003), and thus
change the $M/L$ profile in the N-body simulations

Another potentially important concern is how stellar remnants are
handled. The two components that need to be understood are the neutron
star retention factor and the present-day mass profile for white
dwarfs. The neutron star retention factor appears not to be a serious
issue anymore since the recent N-body simulations from Baumgardt et
al. use very little retention. The contribution from neutron stars to
the central mass profile is likely to be very small. This is not true
of older simulations (e.g., Dull et al. 1997), so one must be aware of
these differences. We know, however, that clusters do contain pulsars,
so the retention factor is larger than zero. The best estimates place
it between 5 and 15\% (Pfahl et al. 2002). Recent results from Heinke
et al. (2005) using Chandra show that 47Tuc's neutron star population
is consistent with a very low retention fraction. A more important
issue is the initial-to-final mass relation for white dwarfs.  The
remnants that have the most influence on the central structure are the
heavy white dwarfs. Nearly all N-body simulations use the stellar
evolutionary models of Hurley, Pols, \& Tout (2000). For the
initial-to-final mass relation, they use an empirical fit to the data
of Jeffries (1997) for 4 white dwarfs in the same cluster. Both recent
observational and theoretical results, however, give a significantly
different relation. The theoretical models of Weidemann (2000) or the
observations of Kalirai et al. (2005) show that the final mass of the
heavy white dwarf is smaller by about 0.2 solar mass compared to
Hurley et al.'s estimate. Kalirai et al. also find a slight, but
noisy, dependence on metallicity, with lower metallicity possibly
producing heavier remnants. For a cluster like G1, the smaller masses
for the white dwarfs will have an important effect on dynamical
friction timescales, and thus we should not expect as large a
contribution to the central mass from them as Baumgardt et
al. suggest. New N-body simulations should include the most up-to-date
initial-to-final mass relation for white dwarfs in order to provide
realistic results.

The final mass of the white dwarfs do not affect our analysis as much
as in the N-body simulations. The orbit-based models rely on having an
estimate of the mass-to-light ($M/L$) ratio in order to provide an
input gravitational potential. The N-body simulations of G1 from
Baumgardt et al. (2003) show that the $M/L$ profile of G1 near the
center is at least constant and possibly even decreasing. Given the
age around 13 Gyr for G1 (Meylan et al. 2001) and a turn-off mass
around 0.9$~\Msun$, massive giant stars tend to dominate the light in
the core; even though there are remnants in the core, the combined
$M/L$ tends to remain constant. The models of Baumgardt et
al. actually show a slight drop in the $M/L$ near the center of G1. In
our analysis we assume a constant $M/L$ ratio. We have tried a variety
of realistic expectations for the $M/L$ variation (as in Gebhardt,
Rich \& Ho) and all models strongly support the existence of a central
black hole.

We now turn to results from newer data and analysis on G1, but argue
that the black hole as presented in Gebhardt, Rich \& Ho is robust.

\section{Data}

One of the larger uncertainties in the previous dynamical models is
the comparison with kinematics at large radii. The overall
mass-to-light ratio is extremely important since it helps to set the
scale at small radii where the influence of the black hole is
seen. Furthermore, orbit-based models require some knowledge of
large-radii kinematics since the models need to constrain the
influence of orbits with highly radial motion. For example, a central
black hole can easily be disguised if one includes a large amount of
radial orbits.  However, in this case, the radial orbits will have a
large effect at large radii, causing a significant drop in the
projected dispersion there.  Thus, by having large-radii kinematics,
one can limit this effect. In our previous analysis, we relied on a
signal-aperture measurement from the ground-based spectroscopy of
Djorgovski et al. (1997) to provide most of the large-radii
leverage. In order to improve upon this, we obtained high
signal-to-noise (S/N) spectra from Keck.

The other issue is that the central light profile was not very well
determined in the previous analysis. The previous HST imaging suffered
from saturation, no dithering, and coarser resolution. We have
obtained improved ({\it HST}) imaging, which we present below.


\vskip 5pt \psfig{file=gebhardt.fig1.ps,width=9cm,angle=0}
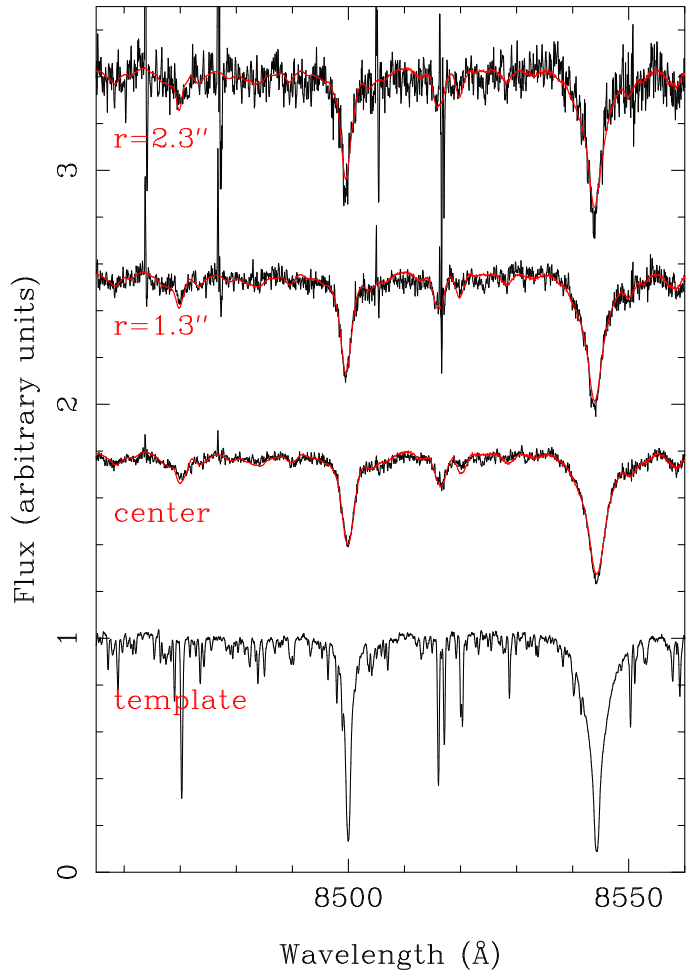
\figcaption{ Keck spectra and fits for G1. The wavelength range
includes two of the Calcium triplet lines (and some weaker lines). The
bottom spectrum is the template used for the estimation of the
velocity profile. The upper spectra come from different radii. The
noisy lines are the data, and the red lines are the template convolved
with the velocity profiles.}

\vskip 10pt


\subsection{Keck Spectra}

The Keck data were obtained on October 19, 2003 using the
high-resolution spectrograph HIRES. We observed a total of 2.8 hours.
We used a 1.72\arcsec\ slit and the 14\arcsec-long D4 decker. The
spectra run from 6390 to 8770 \AA. This setup produces an instrumental
resolving power of R=23,000, or resolution of 13~\kms\ (FWHM) at
8500\AA, which is ideal for the velocity dispersion of G1
(10--30~\kms). The spatial scale in the cross-dispersed direction is
0.191\arcsec\ per unbinned pixel. We binned by two in the cross
dispersed direction since we are mainly concerned about large-radii
kinematics. Around 8500\AA\ the wavelength scale is about 0.0626\AA\
per pixel in the dispersed direction. The FWHM of the seeing during
the observations was about 1.2\arcsec.

We used the reduction package MAKEE (written by Tom Barlow) for the
data analysis. We compared these results to our own reduction
procedure and found no significant differences. However, the ease of
use of MAKEE make it an excellent and ideal package. The MAKEE package
performs standard flattening, traces and extraction. Since we are
interested in spatial information, we extracted spectra along various
radii. With the 14\arcsec\ slit and the size of G1 (half-light radius
around 1.5\arcsec), we always had sky in part of the slit that we used
to subtract from the data. Our furthest radial bin runs from
3\arcsec--4.6\arcsec, leaving us with 2.5\arcsec\ for sky estimation.
We checked whether residual light from G1 contaminates the sky by
running the kinematic extraction with and without the sky
subtraction. While the results with no sky subtraction give
substantially worse results, the kinematics are qualitatively the
same.  Thus, the sky appears to be well measured and subtracted.

Figure 1 presents spectra at three different radii in G1. The bottom
spectrum is a template star. It is clear that the Calcium triplet
lines at all radii are easily resolved with this setup. The S/N in
the central spectrum is 55 per extracted pixel (corresponding to
0.0626\AA\ by 0.382\arcsec). At our largest radii, S/N=6.7 per
extracted pixel (0.0626\AA\ by 1.6\arcsec).  We had five individual
exposures for G1. Each is extracted separately and the continuum
divided out before making combined spectra.

We use the kinematic extraction technique outlined in Gebhardt et
al. (2000c) and Pinkney et al. (2003). This technique provides a
non-parametric estimate of the LOSVD, which is used directly in the
models. Traditionally, one reports the moments of the velocity
profile. Below we calculate the moments, but keep in mind that these
moments are not used in the models but instead the full LOSVD. The
shape of the LOSVD provides important information about the orbital
structure and even the black hole mass; it is important to include the
full velocity profile when making dynamical models.  Table~1 presents
the first four moments of the velocity profile from the ground-based
kinematics. These are the velocity, velocity dispersion, H3 and H4. H3
and H4 represent diviations from a Gaussian using a Gauss-Hermite
polynomial.  H3 is similar to skewness and H4 similar to
kurtosis. Van~der~Marel \& Franx (1993) and Bender et al. (1994)
provide detailed discussion of the Gauss-Hermite expansion. These
first four moments are determined directly from our non-parametric
estimate of the LOSVD. We only present the symmetrized version of the
kinematics. We symmetrize by fitting spectra at the same radii on
opposite sides of the cluster simultaneously to the LOSVD; however,
the LOSVD is appropriately flipped about zero velocity from one side
compared to the other, which is the expected configuration for an
axisymmetric system. The uncertainties in Table~1 have been
symmetrized; our fitting procedure includes the actual uncertainty
distribution, but since this distribution is nearly symmetric we
present only symmetrized uncertainties in the Table. For both the STIS
and Keck data the width of the extraction window varies with radius,
and the uncertainties at each point depends on the surface brightness
and extraction width.

We observed four different template stars: two K4III stars, one K0III,
and one M1III. All templates give nearly identical results for the
kinematics. The rms of the kinematic parameters for the four templates
scatter within the 1 $\sigma$ uncertainties presented in Table 1. Our
final results are based on using just one K4III star but would remain
unchanged with any of the template. The Calcium triplet region is
therefore quite robust to template variations, as discussed in Barth,
Ho, \& Sargent (2002).

We reanalyse the STIS spectra presented in Gebhardt, Rich \& Ho
(2002).  The differences mainly include a different re-sampling and
combining of the individual spectra and a new extraction of the LOSVD.
Given that the instrumental resolution of STIS is approaching the
dispersion of G1 near the center, one needs to take this into account
carefully. Our re-analysis is very similar to our published results,
with the dispersions being about 3\% smaller in the center and 9\%
smaller at the largest radii. These changes are well within the quoted
uncertainties.  The differences at large radii (1\arcsec\ for STIS)
make essentially no difference in the dynamical modeling since they
have large uncertainties, and the high S/N ground-based spectra
dominate the fits there. Thus, the new analysis changes none of the
results but should be used for any future modeling. Table~1 presents
the moments of the LOSVDs for the STIS spectra.


\vskip 5pt \psfig{file=gebhardt.fig2.ps,width=9cm,angle=0}
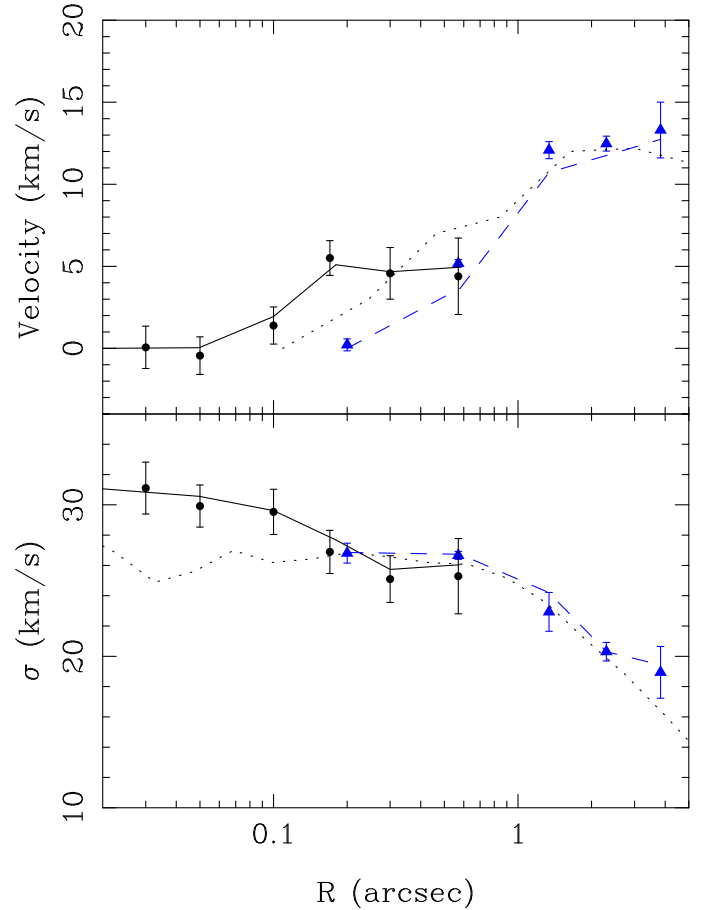
\figcaption{Projected velocity (top) and dispersion (bottom) for G1
including the Keck (blue triangles) and STIS (black circles) data. The
STIS data have been re-analyzed since Gebhardt et al. (2002). The data
have been folded about the center. For both sets of data we plot the
symmetrized version. The solid and dashed lines represents the
kinematics from our best-fit model (including the black hole) for the
STIS and Keck data, respectively. Our model presented here includes
the increase in $M/L$ at large radii. The dotted lines are the
kinematics from Baumgardt et al. (2003b) no-black hole model, which
clearly is low in the central regions (discussed in Section 5).}

\vskip 10pt



\begin{figure*}[t]
\centerline{\psfig{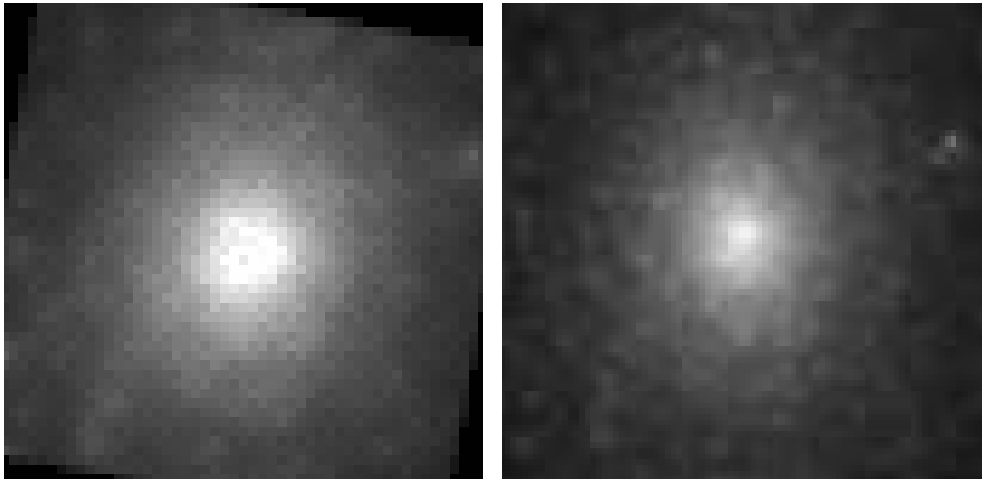}}
\figcaption{HST observations of the central 2.3\arcsec\ diameter using
the PC/F555W on WFPC2 (left) and using HRC/F555W on ACS (right). The
HRC image is deconvolved. The central structure is clearly more
detailed in the HRC image, allowing us to see the effects of
individual bright giant stars.}
\end{figure*}


Figure 2 presents the first two moments (velocity and velocity
dispersion) of the velocity profile for the STIS and Keck spectra.
The agreement between the STIS and Keck data is excellent. At small
radii, the velocity profiles disagree due to seeing in the Keck
data. At large radii, G1 has substantial rotation. At 4\arcsec\,
$v/\sigma$ is around 0.7, which makes it one of the fastest rotators
ever seen for a globular cluster. $\omega$~Cen has the highest
$v/\sigma$ measured for a Galactic globular cluster, with
$v/\sigma=0.3$ (van de Ven et al. 2005).

\subsection{HST/ACS Observations}

We took HST images of G1 with the High-resolution Camera (HRC) on the
ACS. The total integration is 41 minutes in the filter F555W over six
exposures at three positions. Figure 3 shows the central 2.3\arcsec\
with WFPC2 (obtained from the HST archive) and our deconvolved ACS/HRC
image.

The individual exposures are shifted with linear interpolation and
combined with a biweight estimator (Beers, Flynn, \& Gebhardt 1990).
We deconvolve using 140 Lucy-Richardson iterations (Lucy 1974). The
ACS/HRC PSF is taken from the HST handbook and uses Tiny Tim. We tried
different numbers of Lucy-Richardson iterations and anything over 20
produces nearly identical results. The ACS pipeline produces a
drizzled image as well; we find no significant differences between our
image and the drizzled image. For the surface brightness profile, we
determine the central location and scatter of the pixels in ellipses
centered on the cluster. We use an ellipticity constant with
radius. The central location and scatter are determined from the
biweight estimators. The center determination is an important aspect
for the surface brightness profile. In G1, we are spatially resolving
the brightest giants, and bright stars near the center can skew its
determination. Therefore, the center measured from the isophotes at
larger radii provide a more accurate determination. In fact, the
center measured from the outer isophotes is 1.2 HRC pixels
(1.2x0.0266\arcsec=0.03\arcsec) different from the brightest pixel
near the center. If we use the center as determined by the brightest
pixel in the central parts of the image, the surface brightness is
slightly brighter than using our best-measured center. This increase
provides more stellar mass there and makes the black hole mass
determination slightly less significant. In the subsequent analysis,
the $\Delta\chi^2$ between the no-mass black hole and the best fit
changes from 5 to 4.2. However, the bright spot near the center of the
image appears to be a bright star, as the isophotal analysis
suggests. Figure 4 plots the surface brightness profile for G1. We
include our previous determination using WFPC2 images. We can obtain
the exact WFPC2 surface brightness profile by using a center as
defined by the brightest pixel, indicating that our old WFPC2 analysis
was influenced by not being able to resolve the central bright star.
HRC images give us a factor of two improved spatial resolution.


\hskip -10pt \psfig{file=gebhardt.fig4.ps,width=8.5cm,angle=0}
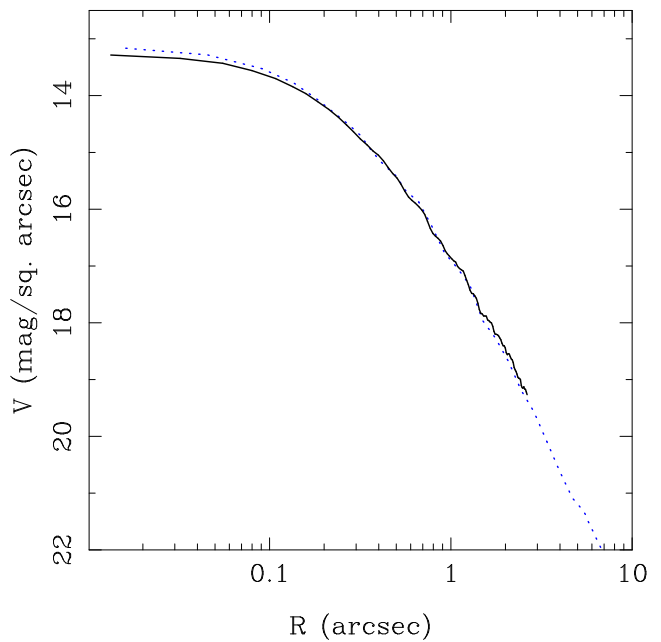
\figcaption{ Surface brightness profiles for G1. The blue dotted line
is from the WFPC2 data that were used in Gebhardt et al. (2002). The
black solid line is from the deconvolved ACS/HRC image, and is used in
the subsequent analysis. The improved spatial resolution allowed us to
better center the cluster and avoid contribution from a bright star,
which causes the decrease in the central brightness seen in the HRC
profile compared to the WFPC2 profile.}



\begin{figure*}[t]
\centerline{\psfig{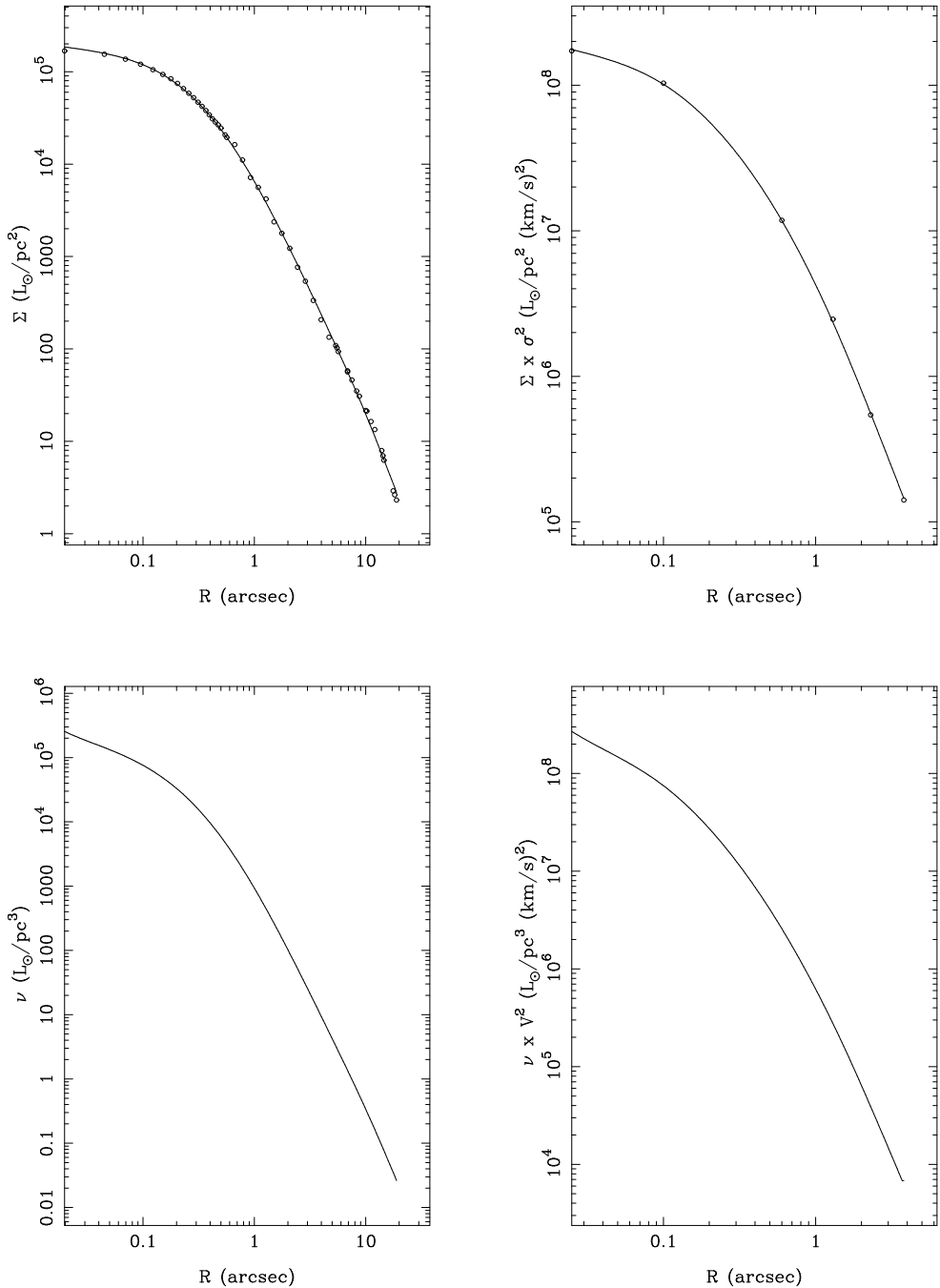}}
\figcaption{The left panels are the projected and internal light
profiles for G1, where we have used a non-parametric deprojection. The
right panels are the projected and internal values for the light
profile multiplied by the second moment of the velocity
distribution. For the projected velocity distribution we are using
$\sigma^2 = v^2+\sigma^2_d$.}
\end{figure*}



\begin{figure*}[t]
\centerline{\psfig{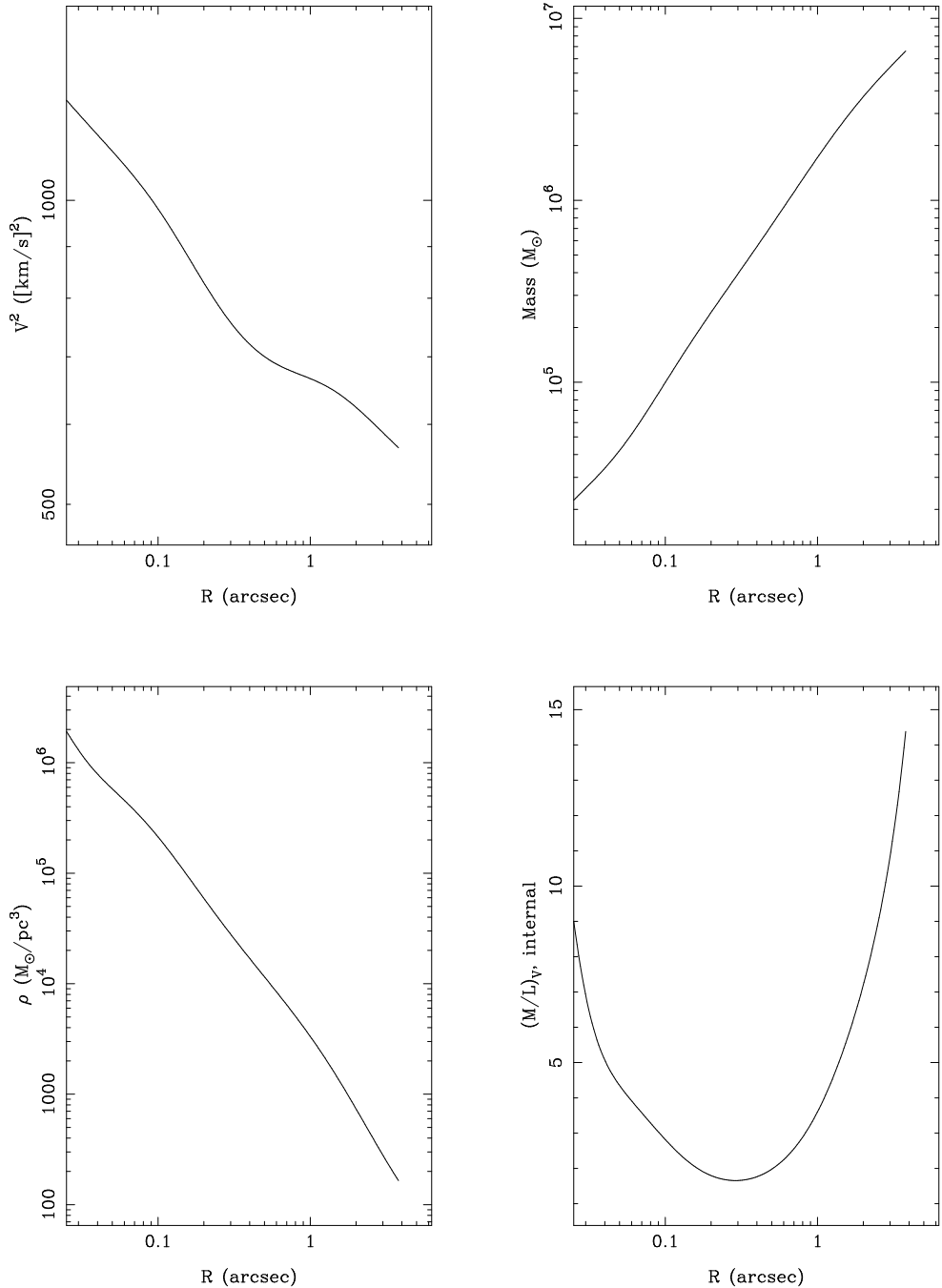}}
\figcaption{ Internal dynamics for G1. The top left is the square of
the second moment of the internal velocity distribution. The top right
is the enclosed mass as a function of radius. The bottom left is the
mass density, and the bottom right is the internal mass-to-light
ratio.}
\end{figure*}


\section{Models}

There are a variety of dynamical models that are available. We model
G1 both with isotropic, non-parametric models, and with fully general
orbit-based models. Baumgardt et al. (2003b) provide a dynamical model
for G1 that is very different from our models. They use evolutionary
models and find the best fit to the data by comparing N-body
simulations with varying initial conditions. Thus, they have a robust
evolutionary model but are limited in their comparison by their
particular N-body runs (i.e., initial conditions, number of particles,
etc.). Our models are static equilibrium models, and we do not follow
the evolution. However, the orbit-based models allow us to modify the
distribution function in order to get the best match to the data while
still being a solution to the Jeans equations. Thus these orbit-based
models have significantly more freedom and produce a more general
result.

However, before discussing the orbit-based models, we model G1 as an
isotropic spherical system. While this is not the case for G1, since
it is flattened with axis ratio 0.75 (as measured from our image), it
does provide a good comparison. Furthermore, the short dynamical
timescales in the central regions of G1 (the central relaxation time
is about $2\times10^8$ years) may cause the orbits to be nearly
isotropic. Since our goal is to constrain the central mass profile and
the rotation is a small component of the velocity dispersion near the
center, we make only a small error by assuming isotropy there. At
large radii, however, this assumption will likely break down due to
the large rotation and long relaxation times.

\subsection{Non-Parametric Models}

We use the non-parametric models as discussed in Gebhardt \& Fisher
(1994). Given a surface brightness profile and a velocity dispersion
profile, the spherical Jeans equation uniquely determines the mass
density profile (and hence the $M/L$ profile) assuming isotropy. This
is a straightforward exercise that we demonstrate in Figures 5 and 6.
Through an Abel deprojection, the surface brightness profile uniquely
determines the luminosity density. Similarly, the surface brightness
times the projected velocity dispersion determines the luminosity
density times the internal (3-D) velocity dispersion. One must use
some degree of smoothing to represent these profiles since the Abel
deprojection involves a derivative.  This is discussed in detail in
Gebhardt et al. (1996), and we using Generalized Cross-Validation
(Wahba 1990) to determine the best-fit smoothing parameters. We have
also tried different smoothings and find very little effect on the
main conclusions.

One then measures the internal velocity by dividing out of the
luminosity density (i.e., dividing the curve in the bottom-left panel
in Figure 5 into the one in the bottom-right panel). The top-left
panel in Figure 6 shows the internal velocity squared versus
radius. It is then straightforward to determine the enclosed mass and
$M/L$ profiles.

The bottom-right panel of Figure 6 shows the $M/L$ profile versus
radius.  The two obvious features are the increase at large radii and
the increase at small radii. The increase at large radii is expected
since that region is dominated by low-mass stars. Indeed, Baumgardt et
al. (2003b) find a similar result in their N-body models. However, the
increase at small radii is not expected from the evolutionary models
with no black hole. The $M/L$ increases by about a factor of 3.5 from
the lowest point. G1 has a turn-off mass around one solar mass. Since
the average mass of the stars in the cluster should be around 0.6
solar masses, the one-solar mass stars, as well as the stellar
remnants of neutron stars and stellar-mass black holes, will sink into
the center by dynamical friction. However, since the turn-off stars
and giants are so bright, they tend to drive the $M/L$ to smaller
values. The stellar remnants drive the $M/L$ to high values, and the
resulting combined $M/L$ profile is to remain relatively flat. In
fact, Baumgardt et al. argue that the central $M/L$ should be slightly
depressed compared to the global $M/L$ profile.

Thus, the increase in the $M/L$ profile at small radii is not
consistent with normal stellar evolution without a central black
hole. We can estimate the mass of the central black hole from the
top-right panel in Figure 6, the enclosed mass profile. In the central
bin (radius=0.025\arcsec) there is $2\times10^4$ solar masses of
material. The contribution from stars and remnants at these radii is
small; we estimate the stellar mass by using the total light inside of
0.025\arcsec, which is about 900 L$_\sun$, multiplied by the $M/L$ of
2.8, giving $0.3\times10^4~\Msun$. Thus, the best estimate for the
black hole mass is about $1.7\times10^4~\Msun$. Uncertainties come
from the noise on the surface brightness and velocity dispersion
profiles, and are around $0.3\times10^4~\Msun$, but these only
represent measurement uncertainties and do not take into account
potential assumption biases.


\vskip 5pt
\hskip -10pt \psfig{file=gebhardt.fig7.ps,width=8cm,angle=0}
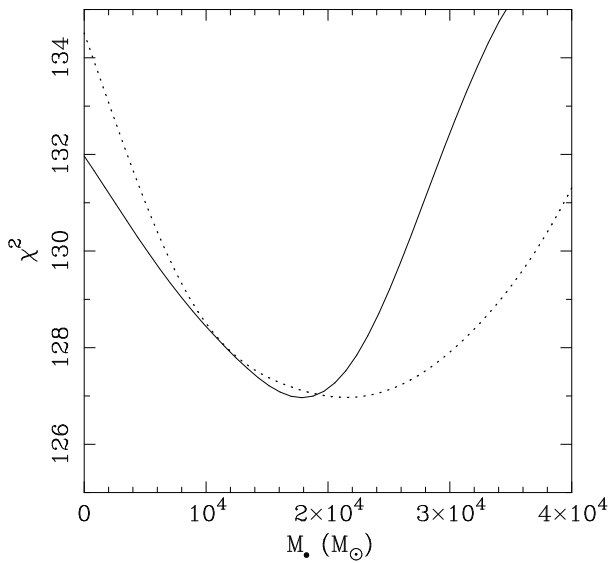
\figcaption{ $\chi^2$ versus black hole mass marginalized over
mass-to-light ratio.  The solid line represents the constant $M/L$
model, while the dashed line has an $M/L$ profile that rises at large
radii according to Figure 6. The difference in $\chi^2$ between the
no-black hole mass and the best fit is 5 for a constant $M/L$, which
implies a significance of over 97\%. The case with a varying $M/L$
shows an even higher $\Delta\chi^2$ with the no-black hole model, and
allows for a higher mass black hole. We add 37 to the $\chi^2$ for the
varying $M/L$ model, since it is a better fit to the data.}

\vskip 10pt


\subsection{Orbit-Based Models}

We use fully general axisymmetric orbit-based models. These are
described in Gebhardt et al. (2000c, 2003), Thomas et al. (2004), and
Richstone et al. (2005).  The models do not rely on a specified form
for the distribution function. Thus, for an axisymmetric system, these
models provide the most general solution. The models require an input
potential, in which we run a set of stellar orbits covering the
available phase space. We find a non-negative set of orbital weights
that best matches both the photometry and kinematics to provide an
overall $\chi^2$ fit. We vary the central black hole mass and refit.

The orbit-based models store the kinematic and photometric results in
both spatial and velocity bins. For G1, we use 12 radial, 4 angular,
and 13 velocity bins. The number of bins is chosen to match the
kinematic extraction windows as well as possible. The data consist of
the seven different STIS positions along a position angle of 25\degr\
up from the major axis and the ground-based data along the major
axis. The point-spread function for both {\it HST}\ and ground-based
observations are included directly into the models, as well as the
slit size. The program matches the luminosity density everywhere
throughout the cluster to better than 0.5\%. The quality of the fit is
determined from the match to the velocity profiles. We use about 4800
orbits to sample the phase space. The orbit number is driven by the
number of spatial bins used in the models (as described in Richstone
et al. 2005). For the binning used for G1, the minimum number of
orbits is around 2400.

Figures 7 and 8 show the results for the models. Figure 7 plots the
one-dimensional $\chi^2$ versus black hole mass, marginalized over
$M/L$ ratio. Figure 8 plots the two-dimensional $\chi^2$ contours as a
function of black hole mass and $M/L$ ratio. For these models we use a
constant $M/L$ ratio. The best-fit black hole mass is
$1.8(\pm0.5)\times10^4\Msun$. The uncertainty represents the span of
$\Delta\chi^2=1$ from the minimum value, which is the 1 $\sigma$ band
for one degree of freedom (since we marginalize over $M/L$).


\vskip 5pt
\hskip -10pt \psfig{file=gebhardt.fig8.ps,width=8cm,angle=0}
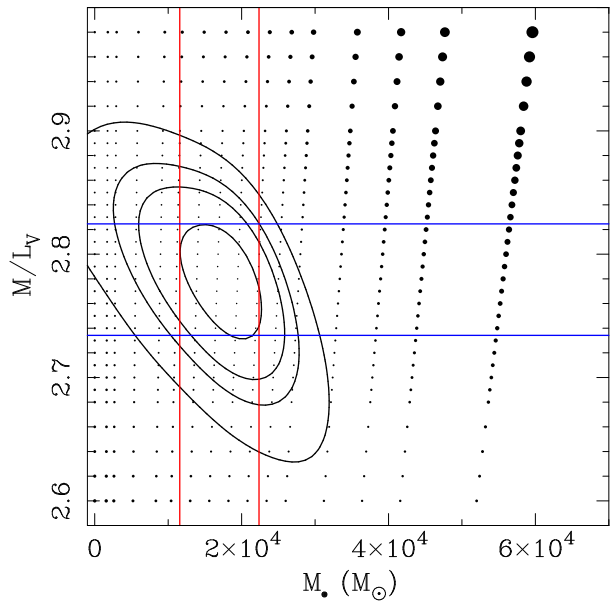
\figcaption{Two-dimensional $\chi^2$ versus black hole mass and
mass-to-light ratio. Each point represents a model, and the size of
the point reflects the value of $\Delta\chi^2$. The contours refer to
$\Delta\chi^2=1.0, 2.71, 4.0, 6.63$, and so corresponds to one degree
of freedom confidence levels of 68, 90, 95, and 99\%. The best-fit
black hole mass is $1.8(\pm0.5)\times10^4\Msun$.}

\vskip 5pt


The contours shown in Figs. 7 and 8 result from a smoothed version of
the $\chi^2$ values. We apply a two-dimensional smoothing spline
(Generalized Cross-Validation: Wahba 1990) to the $\chi^2$
distribution with black hole mass and $M/L$. In this way, we obtain
more realistic uncertainties by interpolating between points with a
smooth function.  The actual $\chi^2$ values lie very close to the
smooth curves seen in Fig. 7, and have insignificant differences for
the best-fit black hole mass. The smoothing is designed to minimize
shot noise in the model values due to grid effects and limited
orbit number.

The $M/L$ ratio of G1 is, however, not constant. At large radii the
$M/L$ ratio increases, but at small radii the stellar $M/L$ ratio is
nearly constant. Thus, we are using an incorrect profile. We include a
varying $M/L$ ratio by having it increase at large radii according to
the profile in Figure 6 (bottom right), but leaving it constant at
small radii (as expected from the simulations of Baumgardt et
al. 2003). The one-dimensional $\chi^2$ is shown in Fig. 7. The
two-dimensional $\chi^2$ is not shown in Fig. 8 but demonstrates the
same result as Fig. 7. With the varying $M/L$ profile, the best-fit
black hole mass is $2.1\pm0.6\times10^4\Msun$. The $\chi^2$ values are
significantly lower---by $\Delta\chi^2=37$---as expected since the
large-radii data show an increase in the dispersion compared to the
constant $M/L$ model.

It is clear by the change in $\chi^2$ that the varying $M/L$ model is
a better fit. However, the $M/L$ that we choose results from an
isotropic analysis and may not be appropriate. Since we are not
exploring a full range of $M/L$ profiles, we choose a conservative
approach for an estimate of the black hole mass by quoting the
constant $M/L$ results. The $M/L$ may increase at large radii due to
anisotropy, a change in the stellar population, or even the inclusion
of dark matter. Globular clusters are thought to have low-mass stars
at large radii due to mass segregation, which is the obvious way in
which the $M/L$ would increase. The increase seen in the isotropic
models is large---around a factor of six---for what is seen in other
globular clusters, so additional explanations from mass segregation
may be in order. We only have kinematics along one axis for G1, so a
full analysis of the $M/L$ profile from our data likely will be
inconclusive. Additional observations of large-radii kinematics of G1
and other massive clusters are required in order to understand the
$M/L$ profile.

\section{Discussion and Conclusions}

Using either a simple non-parametric model or a fully general
axisymmetric model, we find a best-fit black hole mass in G1 of
$1.8(\pm0.5)\times10^4~\Msun$. This mass is consistent with our
previously published result (Gebhardt, Rich \& Ho 2002). This result
is not consistent with the conclusion of Baumgardt et al. (2003b).
There are multiple reason for the differences. First and most
obviously, we are using a significantly improved dataset. The Keck
spectra allow us to determine the $M/L$ significantly better than we
could do before. The improved HST imaging allows us to constrain the
stellar light in the central 0.1\arcsec, which was not possible with
the older WFPC2 data. Both of these data improvements should have an
important consequence on the comparison with Baumgardt et al. In fact,
in Figure 2 we plot the best-fit model of Baumgardt et al. There is a
clear trend for the central dispersions and velocities in the data to
be higher than the values in the model. The increase is consistent
with the black hole mass that we measure. However, this comparison is
not exactly fair since the Baumgardt et al. model was designed to
match the older STIS and ground-based data for G1. It would be
important to compare these newer data to their models again, but it
does appear at this stage that their models with no black hole will
fail to match the data. The other reasons for the difference between
the two groups are our use of the full LOSVD compared to the moments
and the fact that we compare black hole models in a differential
sense.

There are indications that the black hole mass versus velocity
dispersion correlation for galaxies extends down to low masses and
dispersions. Barth, Greene, \& Ho (2005) provide the most recent
dataset that demonstrates this extension. However, their results are
based on estimating the black hole mass from broad-line physics, which
is well-calibrated for higher-mass black holes but not for low-mass
black holes. The black hole mass that we present here for G1 is based
on dynamics, which should be on much more solid footing. In fact,
given G1's integrated velocity dispersion of 25~\kms, the expected
black hole mass from the Tremaine et al. (2002) relation is
$2.3\times10^4~\Msun$. Our measured mass of
$1.8(\pm0.5)\times10^4~\Msun$ implies that the $M_{\bullet}-\sigma$
relation can be extrapolated to these small systems.

G1 may not be a globular cluster but instead the stripped nucleus of a
dwarf galaxy. Meylan et al. (2001) suggest G1 is similar to the
nucleus of NGC~205, and Ferguson et al. (2002, 2005) and Ibata et
al. (2005) report evidence for a disrupted galaxy that is centered on
G1, although there is no evidence for a discrete system. In addition,
Reitzel, Guhathakurta, \& Rich (2004) find no evidence as well for a
separate system from their spectroscopic study of stars in the
vicinity of G1. Nevertheless, compared to globular clusters, G1 would
have the highest measured velocity dispersion (cf. clusters in NGC
5128; Martini \& Ho 2004) and highest v$/\sigma$. The high rotation
seen in G1 may be a further clue to its history since it is difficult
to create such large rotation with an isolated cluster (Baumgardt et
al. 2003b). Thus, it seems plausible that G1 is a stripped
nucleus. There are at least two other dwarf galaxies that have AGN
evidence suggestive of an intermediate-mass black hole. These are
NGC~4395 (Filippenko \& Ho 2003) and POX 52 (Barth et
al. 2004). Further support of dwarf galaxies with black holes come
from the study of Greene \& Ho (2004; see also Barth, Greene, \& Ho
2005). Given our dynamical evidence for a black hole in G1 and that it
may be an accreted dwarf galaxy, there is mounting support for the
picture that some dwarfs contain black holes. Since small galaxies
accrete onto larger galaxies in the hierarchical growth of structure,
there must be observable consequences for having numerous
intermediate-mass black holes sinking into the centers of large
galaxies.

However, it does not appear that all nuclei of small galaxies contain
black holes. M33's nucleus has an upper limit of 1500$\Msun$ (Gebhardt
et al. 2001).  There are no obvious differences between the surface
brightness profiles between G1 and that of the nucleus of M33 (M33 has
a slightly higher central density), but the difference is dramatic in
the velocity dispersion profile; the dispersion of M33's nucleus drops
toward the center while that of G1 rises. It is clear that the data
for G1, compared to that of M33's nucleus, would favor the presence of
a black hole, but it is not obvious why the two systems evidently have
experienced such a different evolutionary history. We clearly need to
address the issue of black holes in small galaxies with a
significantly larger sample.

\acknowledgements

K.G. is greatful for discussions with Tod Lauer about the HST imaging
planning and analysis. We acknowledge grants under HST-GO-09099 and
HST-GO-09767 awarded by the Space Telescope Science Institute, which
is operated by the Association of the Universities for Research in
Astronomy, Inc., for NASA under contract NAS 5-26555. Data presented
herein were obtained at the W.M. Keck Observatory, which is operated
as a scientific partnership among Caltech, the University of
California, and NASA. The Observatory was made possible by the
generous financial support of the W.M. Keck Foundation.  The authors
wish to recognize and acknowledge the very significant cultural role
and reverence that the summit of Mauna Kea has always had within the
indigenous Hawaiian community.

\newcommand{\spa}{\phantom{1}}
\begin{deluxetable}{lrrrr}
\tablecolumns{5}
\tablewidth{0pt}
\tablenum{1}
\tablecaption{Kinematic data for G1}
\tablehead{
\colhead{Radius}   & 
\colhead{Velocity} & 
\colhead{$\sigma$} & 
\colhead{H3}       & 
\colhead{H4}       \\
\colhead{\arcsec}  & 
\colhead{\kms}     & 
\colhead{\kms}     & 
\colhead{}         & 
\colhead{}         }
\startdata
\multicolumn{5}{c}{STIS data} \\
  0.00 & $  0.1\pm1.3$  & $ 31.1\pm1.7$ & $-0.02\pm0.03$ & $-0.07\pm0.02$ \\
  0.05 & $ -0.4\pm1.1$  & $ 29.9\pm1.4$ & $ 0.04\pm0.05$ & $-0.02\pm0.03$ \\
  0.10 & $  1.4\pm1.1$  & $ 29.5\pm1.5$ & $-0.01\pm0.04$ & $-0.02\pm0.02$ \\
  0.17 & $  5.5\pm1.1$  & $ 26.9\pm1.4$ & $ 0.03\pm0.04$ & $-0.09\pm0.01$ \\
  0.30 & $  4.6\pm1.6$  & $ 25.1\pm1.5$ & $ 0.04\pm0.03$ & $-0.05\pm0.01$ \\
  0.57 & $  4.4\pm2.3$  & $ 25.3\pm2.5$ & $-0.01\pm0.05$ & $-0.06\pm0.02$ \\
\multicolumn{5}{c}{Keck/HIRES data} \\
  0.00 & $ 0.2\pm0.4$  & $ 26.8\pm0.7$ & $ 0.04\pm0.03$ & $-0.10\pm0.01$ \\
  0.57 & $ 5.2\pm0.2$  & $ 26.7\pm0.3$ & $ 0.03\pm0.01$ & $-0.07\pm0.01$ \\
  1.34 & $12.1\pm0.5$  & $ 22.9\pm1.3$ & $-0.04\pm0.03$ & $-0.02\pm0.02$ \\
  2.30 & $12.5\pm0.5$  & $ 20.3\pm0.6$ & $-0.10\pm0.02$ & $ 0.00\pm0.01$ \\
  3.83 & $13.3\pm1.7$  & $ 18.9\pm1.7$ & $ 0.13\pm0.06$ & $ 0.09\pm0.03$ \\

\enddata
\end{deluxetable}

\end{document}